# A New Standard DNA Damage (SDD) Data Format

J. Schuemann,[a,1] A. L. McNamara,[a] J. W. Warmenhoven,[b] N. T. Henthorn,[b] K. J. Kirkby,[b] M. J. Merchant,[b,1] S. Ingram,[b] H. Paganetti,[a] K. D. Held,[a] J. Ramos-Mendez,[c] B. Faddegon,[c] J. Perl,[d] D. T. Goodhead,[e] I. Plante,[f] H. Rabus,[g,h] H. Nettelbeck,[g,h] W. Friedland,[h,i] P. Kundrát,[i] A. Ottolenghi,[j] G. Baiocco,[h,j] S. Barbieri,[j] M. Dingfelder,[k] S. Incerti,[l,m] C. Villagrasa,[h,n] M. Bueno,[n] M. A. Bernal,[o] S. Guatelli,[p] D. Sakata,[p] J. M. C. Brown,[q] Z. Francis,[r] I. Kyriakou,[s] N. Lampe,[j] F. Ballarini,[j,t] M. P. Carante,[j,t] M. Davídková,[u] V. Štěpán,[u] X. Jia,[v] F. A. Cucinotta,[w] R. Schulte,[x] R. D. Stewart,[y] D. J. Carlson,[z] S. Galer,[aa] Z. Kuncic,[bb] S. Lacombe,[cc] J. Milligan,[dd] S. H. Cho,[ee] G. Sawakuchi,[ee] T. Inaniwa,[ff] T. Sato,[gg] W. Li,[i,hh] A. V. Solov'yov,[ii] E. Surdutovich,[jj] M. Durante,[kk] K. M. Prise[ll] and S. J. McMahon[ll,1]

[a] Department of Radiation Oncology, Massachusetts General Hospital and Harvard Medical School, Boston, Massachusetts; [b] Division of Cancer Sciences, The University of Manchester, Manchester, United Kingdom; [c] Department of Radiation Oncology, University of California San Francisco, San Francisco, California; [d] SLAC National Accelerator Laboratory, Menlo Park, California; [e] Medical Research Council, Harwell, United Kingdom; [f] KBRwyle, Houston, Texas; [g] Physikalisch-Technische Bundesanstalt (PTB), Braunschweig, Germany; [h] Task Group 6.2 "Computational Micro- and Nanodosimetry", European Radiation Dosimetry Group e.V., Neuherberg, Germany; [i] Institute of Radiation Protection, Helmholtz Zentrum München - German Research Center for Environmental Health, Neuherberg, Germany; [j] Physics Department, University of Pavia, Pavia, Italy; [k] Department of Physics, East Carolina University, Greenville, North Carolina; [l] CNRS, IN2P3, CENBG, UMR 5797, F-33170 Gradignan, France; [m] University of Bordeaux, CENBG, UMR 5797, F-33170 Gradignan, France; [n] Institut de Radioprotection et Sûreté Nucléaire, F-92262 Fontenay aux Roses Cedex, France; [o] Applied Physics Department, Gleb Wataghin Institute of Physics, State University of Campinas, Campinas, SP, Brazil; [p] Centre for Medical Radiation Physics, University of Wollongong, Wollongong, NSW, Australia; [q] Department of Radiation Science and Technology, Delft University of Technology, Delft, The Netherlands; [r] Department of Physics, Faculty of Science, Saint Joseph University, Beirut, Lebanon; [s] Medical Physics Laboratory, University of Ioannina Medical School, Ioannina, Greece; [t] Italian National Institute of Nuclear Physics, Section of Pavia, I-27100 Pavia, Italy; [u] Department of Radiation Dosimetry, Nuclear Physics Institute of the CAS, Řež, Czech Republic; [v] Department of Radiation Oncology, University of Texas Southwestern Medical Center, Dallas, Texas; [w] Health Physics and Diagnostic Sciences, University of Nevada Las Vegas, Las Vegas, Nevada; [x] Division of Biomedical Engineering Sciences, School of Medicine, Loma Linda University, Loma Linda, California; [y] Department of Radiation Oncology, University of Washington, Seattle, Washington; [z] Department of Therapeutic Radiology, Yale University School of Medicine, New Haven, Connecticut; [aa] Medical Radiation Science Group, National Physical Laboratory, Teddington, United Kingdom; [bb] School of Physics, University of Sydney, Sydney, NSW, Australia; [cc] Institut des Sciences Moléculaires d'Orsay (UMR 8214) University Paris-Sud, CNRS, University Paris-Saclay, 91405 Orsay Cedex, France; [dd] Retired; [ee] Department of Radiation Physics and Imaging Physics, The University of Texas MD Anderson Cancer Center, Houston, Texas; [ff] Department of Accelerator and Medical Physics, National Institute of Radiological Sciences, Chiba, Japan; [gg] Japan Atomic Energy Agency, Nuclear Science and Engineering Center, Tokai 319-1196, Japan; [hh] Task Group 7.7 "Internal Micro- and Nanodosimetry", European Radiation Dosimetry Group e.V., Neuherberg, Germany; [ii] MBN Research Center, 60438 Frankfurt am Main, Germany; [jj] Department of Physics, Oakland University, Rochester, Michigan; [kk] GSI Helmholtzzentrum für Schwerionenforschung, Biophysics Department, Darmstadt, Germany; [ll] Centre for Cancer Research and Cell Biology, Queens University Belfast, Belfast, United Kingdom

Schuemann, J., McNamara, A. L., Warmenhoven, J. W., Henthorn, N. T., Kirkby, K. J., Merchant, M. J., Ingram, S., Paganetti, H., Held, K. D., Ramos-Mendez, J., Faddegon, B., Perl, J., Goodhead, D. T., Plante, I., Rabus, H., Nettelbeck, H., Friedland, W., Kundrát, P., Ottolenghi, A., Baiocco, G., Barbieri, S., Dingfelder, M., Incerti, S., Villagrasa, C., Bueno, M., Bernal, M. A., Guatelli, S., Sakata, D., Brown, J. M. C., Francis, Z., Kyriakou, I., Lampe, N., Ballarini, F., Carante, M. P., Davídková, M., Štěpán, V., Jia, X., Cucinotta, F. A., Schulte, R., Stewart, R. D., Carlson, D. J., Galer, S., Kuncic, Z., Lacombe, S., Milligan, J., Cho, S. H., Sawakuchi, G., Inaniwa, T., Sato, T., Li, W., Solov'yov, A. V., Surdutovich, E., Durante, M., Prise, K. M. and McMahon, S. J. A New Standard DNA Damage (SDD) Data Format. *Radiat. Res.* 191, 76–92 (2019).

Our understanding of radiation-induced cellular damage has greatly improved over the past few decades. Despite this progress, there are still many obstacles to fully understand how radiation interacts with biologically relevant cellular components, such as DNA, to cause observable end points such as cell killing. Damage in DNA is identified as a major route of cell killing. One hurdle when modeling biological effects is the difficulty in directly comparing results generated by members of different research groups. Multiple Monte Carlo codes have been developed to simulate damage induction at the DNA scale, while at the same time various groups have developed models that describe DNA repair processes with varying levels of detail. These repair models are intrinsically linked to the damage model employed in their development, making it difficult to disentangle systematic effects in either part of the modeling chain. These modeling chains typically consist of track-structure Monte Carlo simulations of the physical interactions creating direct damages to DNA, followed by simulations of the production and initial reactions of chemical species causing so-called "indirect" damages. After the induction of DNA damage, DNA repair models combine the simulated damage patterns with biological models to determine the biological consequences of the damage. To date, the effect of the environment, such as molecular oxygen (normoxic vs. hypoxic), has been poorly considered. We propose a new standard DNA damage (SDD) data format to unify the interface between the simulation of damage induction in DNA and the biological modeling of DNA repair processes, and introduce the effect of the environment (molecular oxygen or other compounds) as a flexible parameter. Such a standard greatly facilitates inter-model comparisons, providing an ideal environment to tease out model assumptions and identify persistent, underlying mechanisms. Through inter-model comparisons, this unified standard has the potential to greatly advance our understanding of the underlying mechanisms of radiation-induced DNA damage and the resulting observable biological effects when radiation parameters and/or environmental conditions change. © 2019 by Radiation Research Society

## INTRODUCTION

Cellular responses to radiation damage have been studied for many decades, showing the dependency of DNA damage on the delivered dose, the delivery timeframe and the radiation particle type and energy. Numerous models have been developed to explain these responses across a range of end points, including DNA damage, mutations, micronuclei formation, chromosome aberrations and cell survival. Many of these are phenomenological macroscopic models, and simply relate cellular end points to the delivered dose and empirical parameters expressing cell sensitivity, which can depend on the cell line, irradiation conditions and radiation quality. Such phenomenological approaches can capture the overall population-based trends in cell survival that are necessary to describe the effects of radiation therapy, or to estimate effects of exposure to environmental or space radiation. The most common example is the linear quadratic (LQ) cell survival model, which is widely used both experimentally and clinically. To more systematically include the observed dependence of cell survival on the ionization pattern of the radiation modality, i.e., the particle type and energy, various models have been proposed that explicitly include additional physical properties to describe effects relative to a reference radiation. Some models consider the linear energy transfer (LET) (1–4) or other properties related to the structure of the primary irradiating particles and the tracks of surrounding secondary particles (the track structure) in the cell survival calculation, such as the local effect model (5) and the microdosimetric kinetic model (6). The latter two models are used clinically in carbon therapy (7–10). However, these models are also primarily phenomenological and their parameters are dependent on fitting to a selected data set, rather than being based on more fundamental radiobiology.

To advance the field towards more individualized therapies we must study the underlying biological mechanisms of cellular response to radiation and develop effective multi-scale models of radiation action that combine physics, chemistry and biology. Efforts to model cell response have focused on damages to the nuclear DNA, which has long been established as the primary radiation target determining cell viability. The response of cells to radiation has been shown to correlate with the pattern of energy depositions within the nucleus; this correlation is attributed to the resulting differences in patterns and types of DNA damage.

Several decades ago, the first studies using Monte Carlo simulations were performed to link the track structure of different radiation modalities with DNA geometries and the probability of damage induction (11–22). These studies represent the first attempts to apply track-structure Monte Carlo simulations, to mechanistically understand how radiation energy depositions lead to DNA damage. In an ideal scenario, one would use track-structure simulations of the incident radiation to simulate the physical interactions as a means of obtaining nanometer-scale energy depositions and ionizations in accurate geometric models of the cells and their sub-components (nucleus and DNA). After the physical interactions, the resulting radiolysis products and other ionized molecules react in a physicochemical stage, which is followed by migration of the chemical species. At this stage chemical species can react with each other, be scavenged inside the cells or react with components of the cell, such as the DNA. The simulation finishes by determining the direct (caused by physical interactions) and indirect (caused by chemical reactions) damages to DNA. Finally, the DNA damage patterns can be used in mechanistic models of DNA repair kinetics to calculate cell viability, accounting for the damage complexity, along with properties of the cell and the surrounding environment, such as repair deficiencies, cell cycle and oxygenation.

In recent years, several major developments have led to a surge in attempts to mechanistically describe DNA damage and repair kinetics (23–47). An increase in the computational power of standard computers has enabled the simulation of particle tracks in DNA fragments and even whole nuclei (48–53). This has been accompanied by improvements in imaging techniques for studying the responses of cells to ionizing radiation, providing an abundance of data showing the importance of repair pathways and their effect on cell viability. Currently, several Monte Carlo simulation codes exist that can provide the nanoscale track structure of particles passing through a medium, which is typically simulated as water but more recently also includes DNA nucleotide material (23, 54–60). Codes like Geant4-DNA (23, 61), KURBUC (62), PAR-TRAC (60), MC4 (63), RITRACK (57), TRAX (64) and TOPAS-nBio (65) are used by various groups to simulate the track structure of different types of radiation and then score the resulting initial damages to a cell nucleus. One can even go further down in scale and use a molecular dynamics approach such as in the MBN Explorer (66) to model the interactions between molecules. Increasing the simulation detail increases the complexity and makes simulations computationally more expensive. Thus, researchers often apply a multi-scale approach, using macroscopic simulations to capture realistic radiation fields for radiation therapy or mixed-field exposures, and switching to the cell-scale simulations of particle tracks scoring DNA damages in selected cells. This approach can be achieved using a common simulation framework [e.g., Geant4/Geant4-DNA, TOPAS/TOPAS-nBio (67), or the multiscale approach (68–70)], or with the ad hoc coupling of different transport and track-structure codes [e.g., PHITS and PARTRAC (71)].

Each of these codes includes models of DNA structures within the nucleus or cell that are used to obtain the initial patterns of DNA damage. Most of these codes also include the first chemical reactions, i.e., the physicochemical generation of radiolysis products and their subsequent diffusion and interaction (72–77). Thus, these Monte Carlo codes can provide estimates of DNA damages induced both directly (from the initial particle track) and indirectly (from chemical reactions). While these codes frequently differ in their underlying assumptions, have slightly different implementations of particle transport and physics handling and have developed their own data structure and damage pattern definitions, complicating inter-code comparisons of damage induction, most track-structure codes predict reasonably similar yields of double-strand breaks (DSBs), in part because the number of DSBs is often used as a reference dataset.

To fully elucidate the effect of DNA damage induction and repair on cell survival, chromosome aberrations, mutations or other end points of interest, the simulated patterns of damage along the DNA strands, as well as their complexity, must then be combined with models that describe the mechanisms of DNA repair (78). Various groups are working on models to describe these DNA repair kinetics, and to better understand the dependencies of predicted end points on uncertainties and assumptions made in each part of this modeling chain, a direct comparison between models and simulation results from different groups would be immensely useful. However, because of the differences in damage model outputs and dependencies among different damage and repair models, these comparisons are arduous and complex. Repair modeling approaches typically either use an assumed (often random) distribution of damage or are designed specifically to interface with one of the available track-structure codes in an ad hoc fashion (30, 31, 37, 38, 79–83). While typically based on similar principles, these models often employ different approaches and make different assumptions about the underlying repair processes. Inter-comparison between repair models is often complicated by their close links with underlying damage models, which introduces implicit assumptions and dependencies that may not be apparent on simple inspection (84). Providing a common interface would offer much more flexibility and scope to testing different combinations of models, and comparing implicit assumptions and uncertainties.

In addition, there can be important differences among communities in the way DNA damages are defined. Examples include differences in what constitutes a single-strand break (SSB) or a DSB and how they are categorized into different lesion complexities; what factors are considered when describing the nuclear environment; or at what time point the damages are recorded (85). Providing a standard data format that all groups can refer to will help to highlight these differences among the groups and disciplines and provide a platform with which to reconcile them.

Here, we propose such a new "Standard for DNA Damage" (SDD) to facilitate cross-comparisons among the various track-structure Monte Carlo codes and their implementations of first chemical reactions within the cell nucleus, and to link these to mechanistic models of cell repair and the kinetics of DNA damage repair. The proposed standard data format, shown in Fig. 1, provides a new method for cross-code comparisons and promotes collaborations among groups by promoting sharing of DNA damage patterns at selected stages in time, i.e., after the initial energy depositions (direct damages) or after the chemical stage (including indirect damages), as input to calculate the biological end point(s) of interest. By developing a standard data format that various codes can write or read, we provide the means to not only compare the results of different codes and models, but also investigate the influence of each model assumption and cross-validate between models. Testing the dependencies of various observable outcomes on model parameters and their implementation in different models can help us to determine which parts of the models are most sensitive and which parts have only a minor effect on the outcome. In combination with new experimental data of repair processes, in particular with higher temporal or spatial resolution from new microscopy technologies, this can further help to test the models at various stages along the repair process and identify key experiments to advance the field of radiation biology research.

This standard is primarily designed to collect nuclear damage information for eukaryotic cells after radiation damage. However, one can also apply the standard to other sources of DNA damage, e.g., from chemotherapeutic drugs, or to any other organism with DNA, such as bacterial/viral DNA damage (86, 87). In that case, some of the cell-specific information listed in the standard may be omitted. We indicate in some fields where bacterial/viral information can be used instead. While non-nuclear damages can also result in cells becoming nonviable, the proposed standard focuses on the main pathway of cell damage, i.e., damage to the DNA, to provide a compact and easily transferable format.

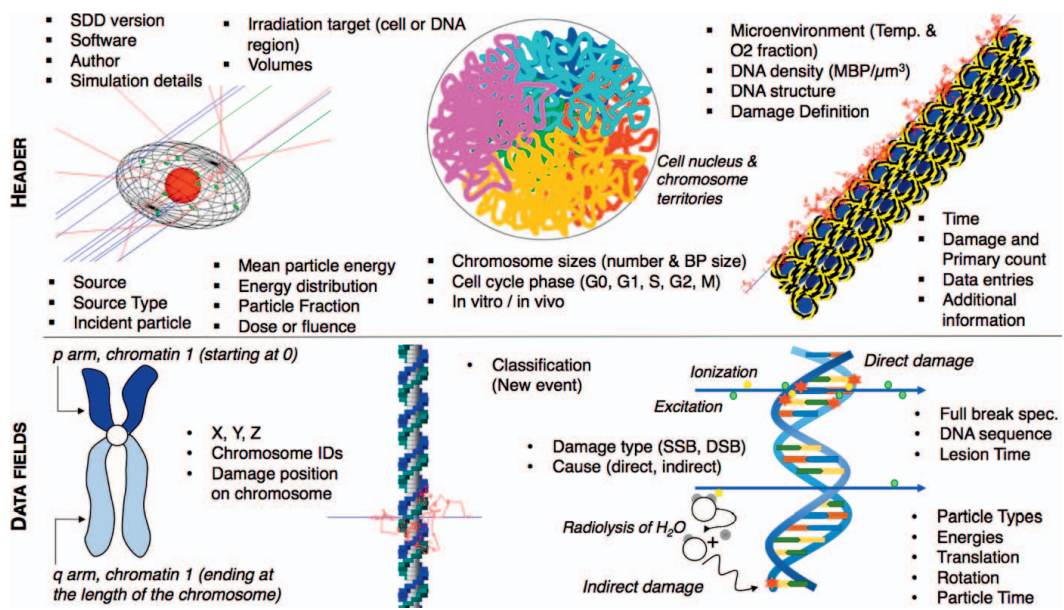

**FIG. 1.** Illustration of the header and data structure of the proposed Standard for DNA Damage (SDD). The information common to all recorded damages is indicated in the header and the information relevant to each damage is indicated in the data section of the SDD file.

## THE NEW STANDARD FOR DNA DAMAGE

The data format for the proposed SDD is based on the format of a typical tuple, i.e., a finite ordered list or sequence of elements. The file format for each damage specification consists of two sections combined in one file that should have the suffix ''.sdd'', for example Filename.sdd. The two sections are:

1. Header Section: A header consisting of a series of factors common to all damage sites in the data block.
2. Data Section: A series of fields defining individual damage sites within the modeled volume.

Our method intends to offer a standard suitable to accommodate a wide range of underlying simulations and DNA repair model designs. To achieve this, the header requires some basic information about the recorded damage patterns for automated read-in of standardized data, while at the same time providing free text sections to expand on the details of the simulation tools. For wide-spread readability, SDD-files employ a comma-separated value format in the header section with each field ending with a semicolon. For the data section, values are separated by a comma or forward-slash and a semicolon is used to indicate a new field. These are the only field separators. Spaces and new lines should be ignored by SDD readers. However, for better readability, we strongly recommend starting a new line for each field.

SDD files are written entirely as plain text (UTF-8 encoding). Due to the variable size of the damage definitions and the sparsity of the data even for radiation exposures of several Gy, a binary format for the data section is considered unnecessary.

### Website and Updates

The SDD data format anticipates that with increasing use cases, numbering schemes will need to be expanded to define additional details or options in some fields. To keep the numbering scheme unique and continue to allow users to share their SDD files without ambiguity, we recommend that requests for new numbering schemes be submitted to the SDD collaboration (represented by the authors of this article, headed by the groups at Massachusetts General Hospital/Harvard Medical School, University of Manchester and Queen's University Belfast) following the steps detailed on the SDD collaboration website: http://standard-for-dna-damage.readthedocs.org/. Each new specification for fields in the header or data block will be assigned a specified number and documentation about all fields will be provided and updated.

### Header Section

The header provides information defining the conditions common to all entries in the Data Section. The structure of the header is presented in Table 1. The header is designed to offer comprehensive information for a large variety of possible damage simulations. A side-effect of this flexibility is that many simulations may not be able to fill all header fields. However, we strongly recommend including all information available. To ensure reproducibility when sharing SDD files, the header contains information on the modeled geometry as well as the irradiation that caused the DNA damage. While the design of the SDD format and the description below is focused on radiation-induced damage, the data structure is flexible enough to allow scoring of other sources of DNA damage, e.g., from chemotherapeutic

**TABLE 1**
**Field by Field Summary of the Header Fields and their Type**

| Field | Value | Notes | Type |
|---|---|---|---|
| 1 | SDD version | Version number of SDD definition | String ("SDDv1.0") |
| 2 | Software | Program name, version and access link if any | String (free text) |
| 3 | Author | Corresponding author, date, references | String (free text) |
| 4 | Simulation details | Description of details of simulation settings and parameters | String (free text) |
| 5 | Source | Description of source properties | String (free text) |
| 6 | Source type | Monoenergetic, distribution, phase space, GCR, … | Int |
| 7 | Incident particles | Definition of primary incident irradiation particle(s) in PDG code format | Int(s) |
| 8 | Mean particle energy | Mean incident energy for each particle in MeV | Float(s) |
| 9 | Energy distribution | Full energy distribution specification | String(s) + Floats |
| 10 | Particle fraction | Fraction of fluence of each particle in field | Float(s) |
| 11 | Dose or fluence | Define dose or fluence in each exposure, or note that the simulation was for a single track | Int + Float (+Float) |
| 12 | Dose rate | Dose rate of irradiation field | Float |
| 13 | Irradiation target | Description of simulated cell or target (DNA) region and microenvironment | String (free text) |
| 14 | Volumes | Shape parameter plus X,Y,Z extents (μm) | 2x (Int + 6 Float) |
| 15 | Chromosome sizes | Number and base pair size of chromosomes | Int + (Int) Floats |
| 16 | DNA density | Density of base pairs in volume (MBP/μm³) | Float |
| 17 | Cell cycle phase | Cell cycle phase index and progression | Int + Float |
| 18 | DNA structure | Additional field to define DNA structure | 2 Ints |
| 19 | *In vitro/in vivo* | Experimental condition | Int |
| 20 | Proliferation status | Proliferative or quiescent and status details | Int + String (free text) |
| 21 | Microenvironment | Temperature (°C) and molar $O_2$ concentration | 2 Float |
| 22 | Damage definition | Define how types of damage were determined | 1 Float + 1 Int + 1 Bool + 2 Float |
| 23 | Time | Time point at which damages were recorded | Float |
| 24 | Damage and primary count | Number of distinct damage lesions scored and primaries simulated | 2 Int |
| 25 | Data entries | Number of fields included in the data section | 14 Bool |
| 26 | Additional information | Field for additional information that may be relevant | String (free text) |
| 27 | ***EndOfHeader*** | Empty field to mark end of header | |

*Notes.* "Int" = parameters that enumerate different possibilities and can take values representing integer numbers between 0 and a maximum value that depends on the field, as detailed in the Header Section. "Bool" = entries that can be either 0 or 1. Entries designated by "Float" are decimal strings to be assigned to floating point variables when reading the SDD file content by a computer program.

drugs. In that case, some of the fields in the header have no value provided and the damage induction may be described in the additional information fields.

Comments can be added to provide additional information on the irradiation, simulation or modeling details. They should be denoted by a block of text, e.g., a new line, starting and ending with #. Any text between these characters should be ignored by the reader codes. Comment lines can be inserted anywhere in the header.

*Description of header fields.* The header consists of 27 fields, each ending with a semicolon. For better readability, each field can be started on a new line, although this is not required *per se*. Table 1 summarizes the proposed fields and their format and additional details for each field are provided below. Each field starts with a string including the "value" tag in the table followed by a comma followed by the types defined in Table 1. If the information for any field is not available, the value string and the field-ending semicolon should still be included in the header. Accordingly, free text sections should not include semicolons except to end the field.

*Field 1, SDD version.* The SDD version number allows tracking future modifications of the file structure and enables automatic transformation of the information in the header and data after such modifications. The version detailed here is SDDv1.0. Thus, the first field should read: "SDD version, SDDv1.0;".

*Field 2, Software.* Here the program name and version number that were used to obtain the DNA damage are described. This can be anything from a simple random sampling function to a combination of dedicated Monte Carlo codes. Due to the free text format, additional information about the software such as an access link (URL) to the software, if available, can be added.

*Field 3, Author.* Here, the corresponding author of the simulations is indicated, to allow for communication about the data provided; the date of the file creation is also indicated. The recommended minimum information is name, email address and date (separated by a comma). Additional references to publications relevant to the simulations can be listed here. The recommended format is: First author (*et al.*), title, journal, edition, page, year,

DOI. If multiple references are included, each reference should be separated by a forward-slash.

*Field 4, Simulation details*: Free text is entered to describe simulation details, ideally providing sufficient information to potentially produce a similar simulation setup. For example, this field should include information about the physics settings, e.g., which secondary particles are included in the simulations with their respective energy cut-off or the names of the cross-section models, where relevant. Also, specifications of the world and transport media corresponding to the interaction cross-section models (e.g., liquid water, vapor water, DNA-like material) can be supplied.

*Field 5, Source*. Free text is entered to describe the particle source used for the simulation. Particularly for scenarios that include multiple-particle irradiations, use phase spaces or other functional forms such as galactic cosmic rays (GCR), this field can be used to add references describing the source, following the structure of field 3, or to provide the URL of a website that defines the source. Additional information relevant to the source that is not covered by the structured data in fields 6–12 should also be added here. Each piece of information should be separated by a comma.

*Field 6, Source type*. Given as an integer, this provides a first overview of the incident particle source, identified as: 1. single or multiple monoenergetic particles; 2. single or multiple particles with energy distributions; 3. a phase space source; or 4. GCRs. For cases 3 and 4, the source should be described in field 5, or users may need to contact the author (field 3) for full source definitions. In addition, for these two options, fields 7–10 may be insufficient and can be left blank; however, if these fields can be used to (roughly) describe the particle distributions, we suggest adding the information. Suggestions for additional options can be submitted to the collaboration website.

*Field 7, Incident particles*. In this field, the radiation type(s) of the incident particle(s) is defined, using the particle specification by the Particle Data Group (PDG) (*88*) to provide flexibility and a comprehensive handling of all known particle types, including (charged) ions and excited states of ions. The radiation source can be an external beam or radionuclides. Each incident particle type can be fully described by a single PDG code (integer). This field lists all incident particle types in the same order that further source definitions should be provided in subsequent fields. Each particle type is separated by a comma. Resulting chemical species, using for example PubChem IDs, are not included at this point, since chemical species typically are created as a result of the irradiation, i.e., from the primary particle.

*Field 8, Mean particle energy*. Here the mean incident particle energy, in units of MeV, is listed as a single float for each particle type listed in field 7 following the same order.

*Field 9, Energy distribution*. This field further specifies the energy distribution of each incident particle in the same order as listed in field 7. For monoenergetic beams (indicated in field 6), this field should take the form "M, 0" for each particle defined in field 7. For other beams, the expected format is a letter specifying distribution ("G" for Gaussian, "B" for bifurcated Gaussian) followed by a comma and a series of forward-slash separated distribution parameters. The mean, μ, is given by field 8. Values required are the variance (G), and left and right variance (B). This field should define one set of parameters for each particle type, using comma separation. For other source formats such user defined distributions or distributions from data tables or for radionuclides, the spectrum should be defined either using these functions or the free text section (field 5), where a link (URL) to a website can be found. Alternatively, users may need to contact the author (field 3), or submit suggestions for additional options of functional forms to be included to the collaboration website.

*Field 10, Particle fraction*. Here the fraction of the fluence, represented by each particle type, is defined, as a single comma-separated number per particle type (defined in field 7, same order).

*Field 11, Dose or fluence*. This field contains, first, an integer specifying whether each field in the data block is for a single-track irradiation (0), a delivered dose (1) or a fluence (2). For the latter two options, the second entry is a float given in Gy for dose, or particles per $\mu m^2$ for fluence. A third value can be added to provide the standard deviation of the mean averaged dose or fluence for multiple exposures. For a single track, the field reduces to "Dose or Fluence, 0;".

*Field 12, Dose rate*. In this field, dose rate in Gy/min is listed. This field provides an easy distinction between space radiation and other low-dose radiation scenarios that can be treated as separate events per incident particle, radiation therapy treatments (in the order of 1 Gy/min), and high-dose-rate deliveries including FLASH therapy and micro-beam or grid therapy (>300 Gy/min) (*44, 89–92*).

*Field 13, Irradiation target*. This field contains free text providing a detailed description of the irradiation target: the cell type, size, cell cycle stage and other properties relevant to the damage induction; size of the nucleus or sub-nuclear region simulated; other geometrical features like mitochondria; and the potential presence of additional factors, such as nanoparticles for radioenhancement or chemotherapeutic drugs. In case of bacterial/viral or mitochondria irradiations, their DNA content can be defined here. Similar to the free text field for the source (field 5), this field should contain information that is not captured by the structured data in fields 14–21.

*Field 14, Volumes*. This field defines the extent of the simulation volume (i.e., the simulated world), and the relevant scoring volume using two sets of comma-separated lists of an integer and six floats. For both volumes, the integer defines the shape of the bounding volume, such as the cell, as either a box (0), an ellipsoid (1) or a cylinder (2); other volume shapes can be added by submitting a request to the SDD collaboration website to extend the SDD by assigning higher value integers. The shape definition is followed by three floats in the order X, Y, then Z, specifying the bounds of the volume in μm. For a box, the values are given in half lengths, i.e., from (+X,+Y,+Z) to (–X,–Y,–Z); for an ellipsoid, the floats define the half axes of the ellipsoid along each of these three axes, i.e., for the special case of a spherical bounding volume, X, Y, Z are identical; for a cylinder, X and Y define the half axes of the ellipsoid along these axes, and Z defines the half length of the cylinder extent (from +Z to –Z). The bounding box thereby also defines the origin of the coordinate system as the center of the bounding box (i.e., the center of the nucleus or cell). The second group of three floats defines Euler rotations, $\phi, \theta, \psi$, respectively, to allow orienting the target in space according to the simulation setup.

The second set of volume definitions (int + 6 floats) follows the same rules as above and defines the scoring volume, e.g., the nucleus. If both volumes are identical, only one has to be defined.

*Field 15, Chromosome sizes.* This field lists the number *N* of chromosomes in the nucleus (or in bacteria/virus), followed by *N* floats for the size of the chromosome in mega base pairs (MBPs). The order of chromosomes listed here should be consistent with the chromosome ID used in field 3 of the data block. Each chromosome should be listed, i.e., a total of 46 for a normal human cell. This allows for the inclusion of cells with missing or multiploid chromosomes. Optionally, if only *N* is provided, the chromosomes are assumed to be uniform in size based on the density stored in field 15.

*Field 16, DNA density.* The field describes the density of the DNA base pairs (BPs) in the scoring volume in units of MBPs per $\mu m^3$ as a single float value. Here the average density over the entire scoring volume is considered, i.e., an average of heterochromatin and euchromatin regions if both are present.

*Field 17, Cell cycle phase.* The field defines the cell cycle and the progression through the phase using an integer and a float. The integer defines the cell cycle numerically as $G_0$ (1), $G_1$ (2), S (3), $G_2$ (4) and M (5). Progression through a phase can be denoted by providing an additional (comma-separated) float with value between 0 and 1. For example, "3, 0.7;" indicates a cell 70% of the way through S phase. This optional float is included to allow more granular inclusion of asynchronous cell populations.

For simulations without a specific cell cycle phase, the value can be set to 0. The cell cycle phase is important to determine the presence of sister chromatids. It further influences the number of chromosome BPs listed in field 14; for cells in (late) S or $G_2$, the number of BPs in a chromosome should only be one half the total number of BPs, as they are repeated and identified by their chromatid number (CR) in field 3 of the data block.

It should be noted that the DNA damage format is designed assuming that each file records responses in a single defined cell type at a particular point in the cell cycle.

However, in *in vitro* or *in vivo* experiments or clinical treatment, the cell population being exposed is typically heterogeneous, with only features such as the composition of the cell population being available, e.g., what fraction of cells are in a given cell cycle phase. Thus, to fully describe biological experiments, it may be necessary to assemble a representative cell population from simulations of different cell cycle stages into a population-level response, with each particular condition stored as an individual SDD file. One can then represent any mixture of cell populations by an adequate assembly of SDD files/scenarios.

*Field 18, DNA structure.* Here, the DNA structure is defined, by two comma-separated integers. The first integer defines the arrangement of DNA as: whole nucleus (0), a heterochromatin region (1), euchromatin region (2), a mixed (heterochromatin and euchromatin) region (3), single DNA fiber (4), DNA wrapped around a single histone (5), DNA plasmid (6) or a simple circular (7) or straight (8) DNA section. Details about higher order DNA assumptions can be added as descriptions in fields 4 or 24, for example by providing the URL of a website or by referring the reader to the author (field 3).

To facilitate cross-code comparisons, options for specially defined geometries are also available. The currently defined reference geometries are a straight DNA section (*100*), a circular DNA plasmid (*101*) and a chromatin fiber (*102*). For exact definitions of these geometries, please refer to the SDD website. The second integer indicates "naked" (0) or wet (1) DNA. Additional values can be added and described by submitting a request to the SDD collaboration website.

*Field 19, In vitro/in vivo.* This field describes the experimental conditions that are simulated. This field is important for both the geometry setup and for considerations of biological response. The condition is defined by two comma-separated integers. The first integer defines if the simulations refer to *in vitro* (0) or *in vivo* (1) conditions. The second integer further explains the conditions; it should be 0 for *in vivo* experiments, (1) for monolayers of cells, (2) for cell suspensions, (3) for 3D-grown tissue models. Additional conditions can be added by submitting requests to the website.

*Field 20, Proliferation status.* This field contains the integer variable to determine the proliferation state of the cell(s) as quiescent (0) or proliferating (1). A second optional string (free text) can be added to describe the status of the scenario, including environmental cues such as serum starvation and innate states like stemness.

*Field 21, Microenvironment.* This field contains two floats. The first value defines the temperature in degrees Celsius, the second the molar oxygen ($O_2$) concentration in the volume in molarity (M). If no values are provided, a standard room temperature of 25°C and normoxic conditions are assumed. Other relevant concentrations such as that of various scavengers should be defined in the free text format of field 13; they are not included due to the wide range of potential scavenging agents. Potential additional fractions or other microenvironment factors can be added by sending a request for expanding the number of parameters given in this field to the SDD collaboration website.

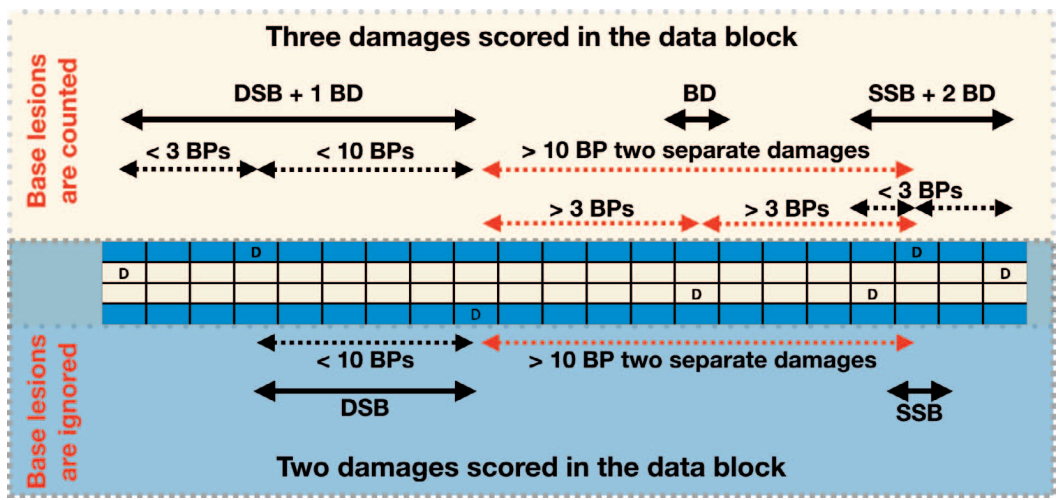

**FIG. 2.** Example of a single DSB recorded with a 10-BP maximum backbone separation. The upper section scores damages if base lesions are counted with a BP separation of up to three BPs as defined in the header field 22.4, scoring three entries in the data block, indicated by the solid arrows. The entries in field 6 would be: a DSB ''1, 2, 1;'' a BD ''1, 0, 0;'' and SSB ''2, 1, 0;'' If base damages are neglected (lower section), the same damage pattern will be scored as two separate damages: a DSB ''0, 2, 1;'' and a SSB ''0, 1, 0;''. The dashed lines demonstrate the separations considered for grouping; red indicates the distances that are larger than the cutoff.

*Field 22, Damage definition.* This field defines how damage is scored and accumulated into distinct damage sites in the data block. It consists of a list of the following values, using comma separation:

1. Integer to define if damages were recorded as those resulting from direct effects only (0) or including chemistry (1). Other types are not currently explicitly included but can be defined by sending a corresponding request to the SDD collaboration website.
2. A Boolean flag to define if the following numbers are listed in number of BPs (0) or in nm (1).
3. This value sets the distance in BPs or nm between backbone lesions that are considered DSBs (float).
4. If this value is set to −1, it indicates that base lesions are not scored. Non-negative values mean that damages to the bases add to the damage complexity and are stored in the data block. In that case, all base damages between backbone damages that form a DSB are stored. This value then determines the distance (in BP or nm) beyond the outer backbone damages where base damages are also stored in the same site (float).
5. Low energy threshold to induce a strand break (or base damage) in eV (float).
6. Optional field to define a linear probability function for damage induction as used in PARTRAC, with the probability $p(E < A) = 0$, $p(E > B) = 1$, and linearly increasing probability from 0 to 1 in the interval from A to B, where A is defined by the 5th value in field 22, and B is given in this field in eV (float). Note: This field will influence the full break specification in the data part of the standard, as demonstrated also in Fig. 2. Fields 22.1, 22.5 and 22.6 influence which interactions are scored as damages, and fields 22.3 and 22.4 determine the distances between and around damages that are clustered in a single break record. However, together with the chromosome position (fields 3 and 4 in the data block), the data block can be post-processed to yield new break clustering using different distances as desired.

An example of field 22 would look like: ''Damage Definition, 0, 0, 10, 3, 17.5;'' translating to: Only counting lesions from direct track interactions, distances are defined in number of BPs, a distance of 10 BP to call two opposite strand SSBs a DSB, base damages are considered, grouping base damages up to 3 BPs on either side of backbone damages in a single site, and only interactions depositing at least 17.5 eV are counted as lesions.

*Field 23, Time.* This specifies the total simulation time for each primary particle, that is, the time from when the source particle was created to the time at which the chemistry simulation ends, i.e., when the damage was recorded, in nanoseconds. For simulations that only consider direct (physics) interactions, this value should be set to 0.

*Field 24, Damage and primary count.* The first integer records the number of distinct damage lesions scored as a single integer and should be identical to the number of fields in the data block divided by the number of fields per damage site (sum of ''true'' values of field 25). The second integer is a counter of how many primary particles were simulated. This value is important to count particles that did not cause any damage to the DNA to accurately represent

TABLE 2
Value by Value Definition of the Data Fields to Score DNA Damages

| Field | Value | Notes | Type | Req? |
|---|---|---|---|---|
| 1 | Classification | Is damage associated with new primary particle or new exposure? And event ID. | 2 Ints | Y,N |
| 2 | X,Y,Z | Spatial X, Y, Z coordinates and extent (μm) | 3x3 Floats | * |
| 3 | Chromosome IDs | ID of chromosome/chromatid where damage occurred and on which arm (long/short) or specification of non-nuclear DNA type. | 4 Int | * |
| 4 | Chromosome position | Location of damage within chromosome | Float | * |
| 5 | Cause | Cause of damage - direct or indirect and number | 3 Int | N |
| 6 | Damage types | Types of damage at site (Base damage, SSB, DSB) | 3 Int | ** |
| 7 | Full break spec | Full description of strand break structure | Special | ** |
| 8 | DNA sequence | DNA Base Sequence around break site | Special | N |
| 9 | Lesion time | Time of each damage induction (in ns) | Special | N |
| 10 | Particle types | PDG list of particles | Int(s) | N |
| 11 | Energies | List of kinetic energies for each particle (in MeV) | Float(s) | N |
| 12 | Translation | Starting position of each particle (in μm) | Floats | N |
| 13 | Direction | Starting direction of each particle (unit vector) | Floats | N |
| 14 | Particle time | Starting time of each particle (in ns) | Floats | N |

*Notes.* Of the fields indicated with ''*'', either field 2 or 3 and 4 are required; similarly, at least one of the ''**'' fields is required. ''Int'' = parameters that enumerate different possibilities and can take values representing integer numbers between 0 and a maximum value that depends on the field, as detailed in the Header Section. Entries designated by ''Float'' are decimal strings to be assigned to floating point variables when reading the SDD file content by a computer program.

the probability of interactions and avoid overestimation of damage induction.

*Field 25, Data entries.* This field contains an array of 14 comma-separated Booleans to indicate which fields of the data block are filled. This field facilitates SDD-reader interfaces.

*Field 26, Additional information.* Allows for additional comments about the simulation that may be relevant for the scored damages. This can, for example, include further details on the physics settings, simulated geometries, material compositions, the source, potential scavenger concentrations in the cell or other descriptions of the simulation or irradiated target that may be helpful to better understand the simulations or improve the biological modeling.

This field can also be used to define new user-specified values for any field in the header or data block. However, we strongly recommend submitting a request to update the standard with such new settings to the SDD collaboration website (http://standard-for-dna-damage.readthedocs.org/) so that the new settings can be officially included in the standard to ensure that all users use the same uniquely defined values.

*Field 27, \*\*\*EndOfHeader\*\*\*.* This is an empty field, used to denote the end of the header and beginning of main text. This field ends the header with: ''\*\*\*EndOf-Header\*\*\*;''.

## Data Section

The data block is recorded in text (UTF-8) format. Each damage site is stored as a group of up to 14 fields, each containing a series of comma- and/or forward-slash-separated fields to define the structure of the damage. Each field will end with a semicolon to indicate the start of the next field in the data block. Some fields are required to identify the original (primary) particle incident on the cell, the position and type of the damage; other fields are optional to provide additional information. Many fields are optional and may be omitted, with the fields used in a particular file defined in the header field 25. We acknowledge that most codes are currently not designed to supply all data fields; we consider the structure also as a motivation for future developments to improve reporting of relevant details. In general, if the information is available in a simulation, it should be added, and all optional fields can be filled to increase the value of the data. The data structure is summarized in Table 2 and detailed below.

*Field 1, Classification.* In this field, the first integer identifies each damage site as a damage from a new radiation exposure, a damage by a new primary particle in the same exposure or another damage from an already recorded primary particle. A new ''exposure'' here means a separate set of the radiation dose or fluence defined in field 11 of the header (e.g., 1.8 Gy) and allows multiple instances of the same irradiation conditions to be recorded in the same file. In that case, the data block contains multiple instances of the same irradiation scheme. For example, for an exposure of 1.8 Gy, a new exposure flag would be set every time the total dose in the target reaches 1.8 Gy, e.g., a total of 42 simulations of 1.8 Gy could be recorded in one file, equivalent to a total simulated dose (in fractions) of 75.6 Gy. Similarly, a new ''event'' refers to damages created by a new (single) primary particle. A new exposure flag also means a new event started. If a particle induces multiple damage sites, e.g., for particle irradiations with high LET, this flag is set to 0 for the second and subsequent damages, indicating damages were caused by the same single particle. The first value of this field is defined by an integer as follows:

0: for a damage caused by the same primary particle as the previous row;

1: for a damage caused by a new primary particle within the same (user defined) exposure; and

2: for a damage which represents the start of a new exposure (which is also necessarily a new primary particle).

The second integer is optional and offers a place to add the event ID, i.e., the number of the primary particle that was simulated, typically counting from 0 or 1 for each exposure.

*Field 2, X, Y, Z (\*).* This field defines the spatial position X, Y, Z of the center and extent of each recorded damage, using coordinates within the bounding box specified by field 14 in the header. The first three value define positions specified as three comma-separated values with unit μm. All subsequent fields are optional but should be included if available. The second set of three comma-separated values defines the maximal position value in X, in Y and in Z and the last three comma-separated values list the minimal values of X, Y, Z, respectively, together defining a box that encompasses the damage. Each 3-tuple of values is separated by a forward-slash; for example, field 2 could read "0.002, 0, 1.2 / 0.004, 0.002, 1.122 / 0.001, –0.001, 1.117;".

*Either field 2 or fields 3 and 4 (Chromosome IDs and Position) must be provided. While both should be listed if possible, the option to define either acknowledges the fact that, depending on the code, not all information may be available.

*Field 3, Chromosome IDs (\*).* This field stores the identity of the chromatid where the damage occurs. The entry consists of four integers. The first integer defines the "DNA structure" as unspecified (0), hetero- (1) or euchromatin (2) regions of nuclear DNA, a free DNA fragment (3) or mitochondrial/bacterial/viral DNA (4). In the case of nuclear DNA, the next three integers are the chromosome and chromatid number and indication of long/short arm. The values are stored, comma-separated, as "CH, CR, CA;" where CH is the chromosome number, CR is the chromatid number and CA is the arm of the chromosome [short (0) or long (1)]. CR is specified as 1 for unduplicated chromosomes, and 1 or 2 to identify the two chromatids in the duplicated chromosome in later S and $G_2$ (and early M) phases. For example, "12, 1, 1;" corresponds to the long arm of chromatid 1 on chromosome 12. Chromosome numbering is assumed to follow the order listed in header field 15. For cells without a specified cell phase or cells in $G_0$ or $G_1$ phase, the chromatid number CR is always 1. In cases where the CA information of short versus long arm is not available, the last number may be left empty.

*Either this value together with field 4 or the X, Y, Z information (field 2) is required.

*Field 4, Chromosome position (\*).* This field indicates the damage position along the chromosome's genetic length. This value is defined as the distance along the chromosome from the start of the short (p) arm towards the end of the long (q) arm. It can be stored either as a value between 0 and 1 (excluding 1) giving the fractional distance along the chromosome at which the break occurs, or, if the value is greater than or equal to 1, as the distance in BPs from the beginning of the short arm (p) to the damage site. In case of non-nuclear DNA, such as DNA fragments or mitochondrial, bacterial or viral DNA, the fraction simply refers to the size of DNA segment provided in the header or, if the value is greater than 1, the BP number along the defined DNA.

*Either this value together with field 3 or the X, Y, Z information (field 2) is required.

*Field 5, Cause (optional).* Offers a flag to identify the cause of the induced damage and a counter for how many damages were caused by direct or indirect events. The first integer classifies the damage type; currently included are options to identify whether the damage is a result of direct physical interactions (0), indirect interactions, i.e., the result of the propagation of any chemical species and after reactions with the DNA (1), caused by a combination of direct and indirect interactions (2), or caused by charge migration (3). Additional options can be included according to the needs of other codes, e.g., to represent damages induced by concomitant drug-based therapies. If additional values of this specifier are needed, a request to update the standard should be submitted to the SDD collaboration website. The second and third integers provide counters for the number of direct and indirect damages at the site, respectively.

Additional information about the damages, e.g., which damage was induced by which process, and more specification of the indirect damages (e.g., fixation by OH, stabilization of $R°$ by $O_2$ or $O_2^-$) can be recorded in field 7.

*Field 6, Damage types (\*\*).* This field provides a high-level specification of the type of damage present at a given site in terms of base damages, backbone (single strand) and DSBs (defined as exactly two single-strand damages within the separations defined in the header), or a combination of these. This classification can be seen as a numerical description of many other damage classification metrics (*21, 93*), effectively grouping these damages into broader categories according to the expected biological severity of the damage. Damages separated by less than the minimum distance of BPs defined by the damage definition in the header (field 22.4) are scored in a single data block, i.e., they are considered to be a single cluster of damages. Repair codes can either convert these clusters to a lesion or use the information in field 7 (if provided) to define lesions. An example of how lesions are grouped based on the information provided by field 22 in the header is shown in Fig. 2.

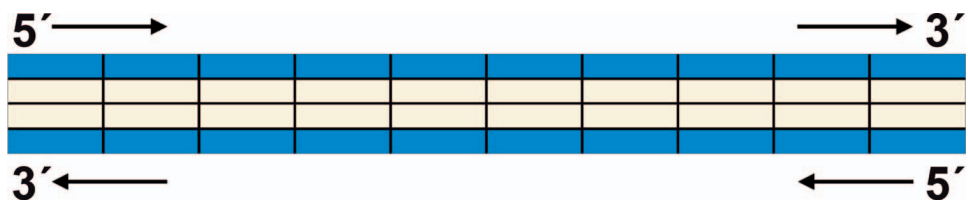

FIG. 3. Structural design of the detailed damage scoring in field 7 of the data structure.

The damages are stored as three comma-separated integers: the first integer lists the number of base damages; the second integer is the total number of single backbone breaks, including those contributing to the formation of a DSB; and the final number is a binary (0 or 1) indicating the presence of a DSB, i.e., if lesions occurred on both backbones within the BP range defined in the header. For example, "3, 2, 1;" would represent a damaged DNA site consisting of three (3) base damages with two (2) backbone damages that are on opposing strands within the BP limit and thus are counted as a DSB (1). Additional examples are listed in Fig. 4.

\*\*Either field 6 or 7 are mandatory, but if available, the full damage structure should always be included, as discussed below, to provide more details of the break structure. This field is intended to provide a high-level summary and support models that do not calculate the full structure of individual breaks and rather rely on numbers of SSBs and DSBs and their distribution.

*Field 7, Full break spec (\*\*).* This field allows for a full specification of the structure of the damage. We apply a four-strand structure, using a $4 \times N$ array, with the rows consisting of the backbone (row 1) and bases (row 2) of the 5′ to 3′ strand, and the bases (row 3) and backbone (row 4) of the 3′ to 5′ strand. The base positions (columns) are aligned reading from the short arm (p) towards the long arm (q), beginning with base 1, which is defined as the first involved alteration at the position corresponding to field 4 (if provided). Thus, increasing columns for strand 1 correspond to the 5′ to 3′ direction and the 3′ to 5′ on strand 2. The design of the data structure is shown in Fig. 3, with blue fields corresponding to the backbones and light orange to the base pair fields.

All unmarked sites are assumed to be unaffected, while damaged sites (strand or base damages) are marked numerically.

Strand damages are indicated as 1 for point breaks on a strand from direct effects, 2 for lesions (losses or attachments) from a single indirect damage, 3 for multiple damages to the same base pair strand from two or more direct or indirect interactions. In addition, 0 can be used to indicate nondamage-inducing interactions, i.e., events that are below the damage induction threshold. Also, 4+ can be used for adducts or other modifications, according to the source program, e.g., to define additional details such as the type of indirect reaction that occurred. This may be relevant to account for differences in repair likelihood for different types of reactions from radiolysis such as dehydrogenation, OH addition or deoxyribose damage. All additional numbering schemes should be submitted to and detailed on the collaboration webpage to ensure a unique numbering scheme.

Base damages are indicated in the same manner as strand breaks, with 1 for point losses from direct effects, 2 for lesions (losses or attachments) from indirect effects, and 3 for multiple damages from direct or indirect damages. 0 can be used to indicate nondamage-inducing interactions. Again, all additional suggested numbering schemes should be submitted to and detailed on the collaboration webpage to ensure a unique numbering scheme.

The damages are recorded in a comma-separated list of: Strand (row), Base pair (column), Damage type, with each damage separated by the delimiter "/". Damage events are recorded by row, beginning with the 5′ to 3′ backbone, then its accompanying bases, then the 3′ to 5′ bases, then finally their backbone. Within each row, damages are then recorded in the 5′ to 3′ direction.

To illustrate the syntax of this field, example break types are shown in Fig. 4, ranging from a single base deletion to a highly complex damage with multiple losses in both bases and strands. These are presented both as schematic DNA sections (Fig. 3), and as an accompanying structure definition below each image in Fig. 4.

For example, for the case of multiple base DSB with overhang, with both direct and indirect damages shown in Fig. 4D, the definition reads: "1, 2, 1 / 1, 3, 1 / 1, 4, 1 / 1, 5, 1 / 1, 9, 0 / 2, 1, 1 / 2, 2, 1 / 2, 3, 2 / 2, 4, 2 / 2, 5, 3 / 3, 3, 1 / 3, 4, 2 / 3, 5, 1 / 3, 6, 2 / 3, 7, 0 / 4, 3, 2 / 4, 4, 2 / 4, 5, 2 / 4, 6, 1;". The BP count of each damage site starts with the first occurring damage, which in this case is on the 5′ to 3′ strand base (2, 1, 1). However, the damages are stored starting with the 5′ to 3′ strand backbone, and the first damage for this strand occurs on the second BP, so the damage definition starts with (1, 2, 1). The first 1 indicates the strand, the middle 2 indicates that a damage occurred on the second BP, and the 1 for the last value shows that this was a direct damage. The second, third and fourth triplets (1, 3, 1 / 1, 4, 1 / 1, 5, 1) similarly define three more direct damages on the following three BPs. The fifth triplet (1, 9, 0) shows that there was another interaction with the strand that did not result in a lesion. If additional damages on this strand would have occurred within this break, it would be listed next, for example an additional single direct backbone damage at BP 10 would add a (1, 10, 1) next.

**A** Single base deletion from direct damage:

D

Result: Field 3: 1, 0, 0;
Field 7: 2, 1, 1;

**B** Single base deletion from direct damage and single strand break from indirect damage, plus two non-damaging events:

D                    I
                                        *
                                                              *

Result: Field 3: 1, 1, 0;
Field 7: 1, 3, 2 / 2, 1, 1 / 3, 6, 0 / 4, 8, 0;

**C** Point double strand break, all direct damage:

D
D
D
D

Result: Field 3: 2, 2, 1;
Field 7: 1, 1, 1 / 2, 1, 1 / 3, 1, 1 / 4, 1, 1;

**D** Multiple base double strand break with overhang, with direct, indirect, multiple and non-damages:

        D       D       D       D                       *
D       D       I       I       M
                D       I       D       I       *
                        I       I       D

Result: Field 3: 9, 8, 1;
Field 7: 1, 2, 1 / 1, 3, 1 / 1, 4, 1 / 1, 5, 1 / 1, 9, 0 / 2, 1, 1 / 2, 2, 1 / 2, 3, 2 / 2, 4, 2 / 2, 5, 3 /
3, 3, 1 / 3, 4, 2 / 3, 5, 1 / 3, 6, 2 / 3, 7, 0 / 4, 3, 2 / 4, 4, 2 / 4, 5, 2 / 4, 6, 1;

**E** Highly complex damage, with both direct and indirect damages and multiple damaged sites:

                                I       M
        I               I       D       D       D
                        D       M       I               D
D                       D       I       D

Result: Field 3: 10, 9, 1;
Field 7: 1, 4, 2 / 1, 5, 2 / 1, 6, 3 / 1, 7, 2 / 1, 8, 1 / 2, 2, 2 / 2, 4, 2 / 2, 5, 1 / 2, 6, 1 / 2, 7, 1 /
3, 4, 1 / 3, 5, 3 / 3, 6, 2 / 3, 7, 2 / 3, 9, 1 / 4, 1, 1 / 4, 4, 1 / 4, 5, 2 / 4, 6, 1;

**FIG. 4.** Examples of the damage definition structure used in field 7 of the data structure. "*" = interactions that were not sufficient to cause a damage, i.e., below the cut-off defined in the header; "D" = direct damages; "I" = indirect damages; "M" = multiple damages from any combination of D and I events. These events are defined by values of 0, 1, 2, 3, respectively.

The group "2, 1, 1 / 2, 2, 1" likewise defines a block of damages on the 5′ to 3′ bases, starting at BP 1 up to BP 2 caused by direct damages, and the next "2, 3, 2 / 2, 4, 2" indicates that the next two bases were damaged from indirect processes. "2, 5, 3" shows that the fifth BP was hit by multiple events in any combination of direct and indirect lesions. The opposite bases start with 3 and the block "3, 3, 1 / 3, 4, 2 / 3, 5, 1 / 3, 6, 2" defines damages on BP 3–6 alternating between direct and indirect damages, followed by another interaction that did not result in a lesion "3, 7, 0". The 3′ to 5′ backbone has the same positions damaged but with the first three damages from indirect processes recorded as "4, 3, 2 / 4, 4, 2 / 4, 5, 2", and an additional direct damage "4, 6, 1".

**Either the full break spec or field 6 (damage types) needs to be included, and ideally both will be provided.

While this field can be omitted if field 6 is provided, all codes simulating the induction of DNA should strive to eventually provide the full break specification. This field, in combination with fields 3 and 4 (chromosome position) can be used to identify exactly where along the strand the damage occurred, to obtain the number of non-hit BPs on either side of the lesion.

*Field 8, DNA sequence (optional).* This field is provided to further specify the surrounding DNA sequence of the site that was damaged. Information about the structural geometry (e.g., heterochromatin or euchromatin) should be provided in field 3. This field consists of a $M \times N$ array to record the DNA sequence. Here, $M$ refers to the number of strands involved, i.e., 2 for a standard double-helix DNA section. $N$ is the number of BPs involved in the damage definition; for the example in Fig. 4C it would be 1, for Fig. 4E it would be 9. For a double helix without mismatched or modified bases, the array can also be reduced to a 1 × N array without loss of information.

For models incorporating the actual physical structure of individual bases, the DNA sequence along the strand can be included in this field. The design uses the same layout as the break structures above, beginning from the 5′ end at the position of the first damage. Bases are written in sequence for the 5′ to 3′ strand, stored as strings of integer values, with bases denoted as: Missing=0, A=1, C=2, T=3, G=4. Backbones are not specified in these data.

This field is kept optional since most codes do not yet consider the DNA sequence. However, evidence exists that specific types of individual lesions formed by ionizing radiation differ for A, G, T and C (94). There is also evidence in the literature that the larger-scale base sequence can have an effect on the types and quantities of individual lesions created by irradiation, and on how this is repaired (95, 96). An example of this latter effect would be hole migration along a DNA molecule. These effects are generally not yet considered in most current Monte Carlo DNA damage models, although the RADAMOL code (75) offers an option to include charge migration. Such migrations have potentially important implications for modeling of DNA repair.

*Field 9, Lesion time (optional).* This field is provided to add a time for each induced damage in nanoseconds starting from the first recorded damage, using the same order as in field 7. If only a single value is given here, that is assumed to be the time at which the whole damage site enters the simulation. The values are recorded, separated by ''/''. For example, for the case shown in Fig. 4B denoted as (1, 3, 2 / 2, 1, 1 / 3, 6, 0 / 4, 8, 0;), the time structure could be ''2.1 / 0 / 0.0000008 / 0.000001''. This translates to an event with an initial direct damage (on 2, 1, 1), a direct base and a direct backbone damage below the break threshold 0.8 and 1 fs later (3, 6, 0) and (4, 8, 0), respectively, and an indirect damage 2.1 ns after the first break (1, 3, 2).

The remaining fields, discussed below, are optional fields to describe the primary particles from the irradiation source that actually caused the recorded damage by itself, through secondary particles or chemical reactions. The primaries should be defined in the Header Section.

*Field 10, Particle types (optional).* Field 10 defines the particle type for each involved source particle using comma-separated integers (PDG values). For single-particle-type irradiations, this field is already defined by header field 7 and can be omitted.

*Field 11, Energies (optional).* This field contains corresponding initial particle energies of the particles defined above, in MeV, one comma-separated entry per particle defined in field 10. Similar to field 10, for monoenergetic irradiations, this field is already defined in header field 8.

*Field 12, Translation (optional).* This field contains the three vectors (X, Y, Z) of the starting points of the particles, in μm, relative to the center of the world volume, stored as a comma-separated list of one 3-tuple for each particle, with each value within the 3-tuple separated using ''/''.

*Field 13, Direction (optional).* This field contains Euler rotation angles (φ, θ, ψ) for the above particles following the same style as for the translations to define their direction. A rotation of 0/0/0 is defined as having the particles propagate along the +Z direction.

*Field 14, Particle time (optional).* This field contains a list of comma-separated floats giving the start time in ns of each particle defined in field 10 with t = 0 defined as the time when a new exposure starts (see field 1 and header field 11). This may be particularly important for very low-dose-rate exposures such as those from GCRs.

### Dissemination and Repository

We have set up a website (https://standard-for-dna-damage.readthedocs.org/), where the standard is documented and to which new requests for enumeration schemes can be submitted. In addition, we provide a link to our GitHub repository, which offers selected example codes and provides a place to share SDD files. Such a repository will be useful for modelers who do not have the resources to perform their own full damage simulations, and to test (new) damage models against other (published) models that provided data here. This repository also includes a code to generate random damage distributions in the SDD format and example SDD files. (See Appendix for further details).

### DISCUSSION

The outlined standardized data format for DNA damages (SDD) is intended to provide the basis for cross-disciplinary investigations of DNA damage induction and ensuing kinetics of DNA repair mechanisms. By standardizing the recording format of the distribution of damages and their structural pattern for single cells and nuclei, we anticipate creating synergies among various developments in modeling cellular response to DNA damage.

The standard has been developed in anticipation of several future developments in the field. For example, modeling of the initial chemical reactions is becoming increasingly common, allowing for inclusion of more sophisticated effects such as the potential production of additional reactive species induced by shock waves from high-LET ions (97) or neutralization effects of Auger emitters (98). With an increased range of chemical reactions simulated, repair mechanisms and accuracy can be adjusted depending on the particular indirect effect classes of DNA damage. Similarly, radiation therapy is often only one of the delivered treatment modalities, combined with chemotherapy or immunotherapy. Modeling of multi-modality treatments is an emerging field, and good-quality input data from each modality is essential. The effects of other treatments can be partially included in the SDD by adding DNA damages from drugs. For therapeutics that inhibit certain repair pathways the biological models will have to be adjusted, with the SDD providing detailed DNA damage maps.

While the nucleus is the primary target in radiation therapy, the standard is flexible enough to also be used to describe damages to mitochondrial DNA or DNA in viruses or bacteria (in a separate file). However, for these cases many of the optional fields in the SDD may not be relevant. The SDD has been designed to allow a high level of flexibility. Many of the entries are optional and are only included to encourage the user to think about the concepts and, if possible, include these details as they may become useful for repair kinetics.

Overall, we anticipate that the SDD data format will greatly reduce the burden of sharing analysis tools and thus, facilitate the formation of new collaborations. Using standardized data will allow researchers to test the predictions from different models simply by feeding the SDD data to another code. The standard already is (or will soon be) supported by the following codes: DaMaRiS (*39*), gMicroMC, MC4 (*63*), MCDS (*27, 79*), PARTRAC (*60*), PHITS (*99*), RADAMOL (*75, 100*), RITRACK (*57*) and TOPAS-nBio (*65*), as well as by users of Geant4-DNA (*23, 61*). By providing a clearly defined standard and example codes of scorers for some of the models, we hope this encourages other existing and newly developed codes to offer interfaces to the SDD data format for use as a scorer or as damage distribution input for repair models.

## CONCLUSION

We have developed a new Standard DNA Damage data format. The SDD has been designed to interface at the point where physics and chemistry simulations, at the DNA scale, meet biological modeling efforts. With this standard, we hope to provide modelers with a new tool to test their model design and dependencies on underlying physics properties. In combination with the supported collaboration website, the SDD offers access to the most accurate available damage simulations and provides a platform for inter-code comparisons of the underlying track-structure Monte Carlo simulation codes and their assumptions in the description of physics, chemistry and the geometrical arrangement of DNA. This standard will play a significant role in advancing our understanding of DNA response to radiation insults by creating the basis for a wide-spread interdisciplinary collaborative effort.

## APPENDIX

### Example SDD Files

With many detailed options of scoring the DNA damage in the SDD format, we believe it is helpful to illustrate the format with example SDD files. We have created two files using the McMahon Empirical Model version 0.3 to generate DNA damages for an irradiation of a cell to 1 Gy with a 0.975-MeV monoenergetic proton beam. The two files use identical irradiation setups, but the first (DNA Damage Proton0.975 MeV 1 Gy full.txt) fills all available scoring blocks (see header field "Data entries"), including records of the full damage definition of data field 7. The second file (DNA Damage Proton0.975 MeV 1 Gy minimal.txt) showcases an SDD file that only fills a minimum of 3 of the 14 data sections. Nevertheless, even the minimal data format offers useful information about the frequencies of various damage types. Links to additional examples and example codes to produce SDD files can be found on the SDD website (http://standard-for-dna-damage.readthedocs.org).

## ACKNOWLEDGMENTS

We acknowledge support from the STFC-funded Global Challenge Network+ in Advanced Radiotherapy Multi-Scale Monte Carlo Modeling for Radiotherapy Sandpit (no. ST/N002423/1 to JS, ALM, JWW, NTH, KK, MJM and SJM). We also acknowledge support from the NIH/NCI [grant no. R01CA187003 ("TOPAS-nBio: a Monte Carlo tool for radiation biology research") to JS] and NASA contract NNJ15HK11 to IP.

Received: August 14, 2018; accepted: October 9, 2018; published online: November 8, 2018